\documentclass[journal]{IEEEtran}
\pdfoutput=1

\usepackage{ifpdf}
\usepackage{subfigure}
\usepackage{graphicx}
\DeclareGraphicsExtensions{.pdf,.jpg,.png}
\usepackage[cmex10]{amsmath}
\usepackage{amssymb}
\usepackage{array}
\usepackage{mdwmath}
\usepackage{mdwtab}
\usepackage{pslatex}
\usepackage{xcolor}
\usepackage{textcomp}
\usepackage{placeins}
\usepackage{eqparbox}
\usepackage{url}
\usepackage{stfloats}
\usepackage{multicol}
\usepackage{subfigure}
\usepackage{float}
\usepackage{algorithmicx}
\usepackage{algpseudocode}
\usepackage{amsthm}
\usepackage{comment}
\usepackage{gensymb}
\usepackage{balance}

\begin{document}

\title{Dense Moving Fog for Intelligent IoT: Key~Challenges and Opportunities}

\author{Sergey~Andreev, Vitaly~Petrov, Kaibin~Huang, Maria~A.~Lema, and Mischa~Dohler
\thanks{This work was supported by the Academy of Finland (Project PRISMA). V. Petrov acknowledges the support of the HPY Research Foundation by Elisa.}
\thanks{S.~Andreev is with Tampere University of Technology, Finland and King's College London, UK.}
\thanks{V.~Petrov is with Tampere University of Technology, Finland.}
\thanks{K.~Huang is with The University of Hong Kong, Hong Kong.}
\thanks{M.~A.~Lema and M.~Dohler are with King's College London, UK.}
\thanks{\copyright\,\,2018 IEEE. The work has been accepted for publication in IEEE Communications Magazine, 2019. Personal use of this material is permitted. Permission from IEEE must be obtained for all other uses, including reprinting/republishing this material for advertising or promotional purposes, collecting new collected works for resale or redistribution to servers or lists, or reuse of any copyrighted component of this work in other works.}
}

\maketitle

\begin{abstract}
As the ratification of 5G New Radio technology is being completed, enabling network architectures are expected to undertake a matching effort. Conventional cloud and edge computing paradigms may thus become insufficient in supporting the increasingly stringent operating requirements of \emph{intelligent~Internet-of-Things (IoT) devices} that can move unpredictably and at high speeds. Complementing these, the concept of fog emerges to deploy cooperative cloud-like functions in the immediate vicinity of various moving devices, such as connected and autonomous vehicles, on the road and in the air. Envisioning gradual evolution of these infrastructures toward the increasingly denser geographical distribution of fog functionality, we in this work put forward the vision of dense moving fog for intelligent IoT applications. To this aim, we review the recent powerful enablers, outline the main challenges and opportunities, and corroborate the performance benefits of collaborative dense fog operation in a characteristic use case featuring a connected fleet of autonomous vehicles. 
\end{abstract}

\section{Advent of Intelligent IoT Applications}
\label{sec:intro}

In December 2017, the 3rd Generation Partnership Project (3GPP) has approved the first implementable part of a global \textcolor{black}{fifth-generation} (5G) standard in the form of the Non-standalone (NSA) 5G New Radio (NR) specification. Being a part of their Release 15 \textcolor{black}{delivered in June 2018}, the NSA 5G NR leverages the existing LTE network as a control and signaling `anchor', while the 5G NR air interface acts as a data pipe. Another part of the latest Release 15 is the Standalone (SA) 5G NR aiming to utilize the next-generation network architecture for both the control and the user plane, which carries the actual data. While NSA 5G NR is more suitable for augmenting enhanced mobile broadband services as it builds upon the existing \textcolor{black}{fourth-generation Long-Term Evolution} (4G LTE) infrastructure, the SA 5G NR becomes a true enabler of ultra-reliable low latency communication~\cite{7397856}.

\begin{figure}[!ht]
    \centering
    \includegraphics[width=0.8\columnwidth]{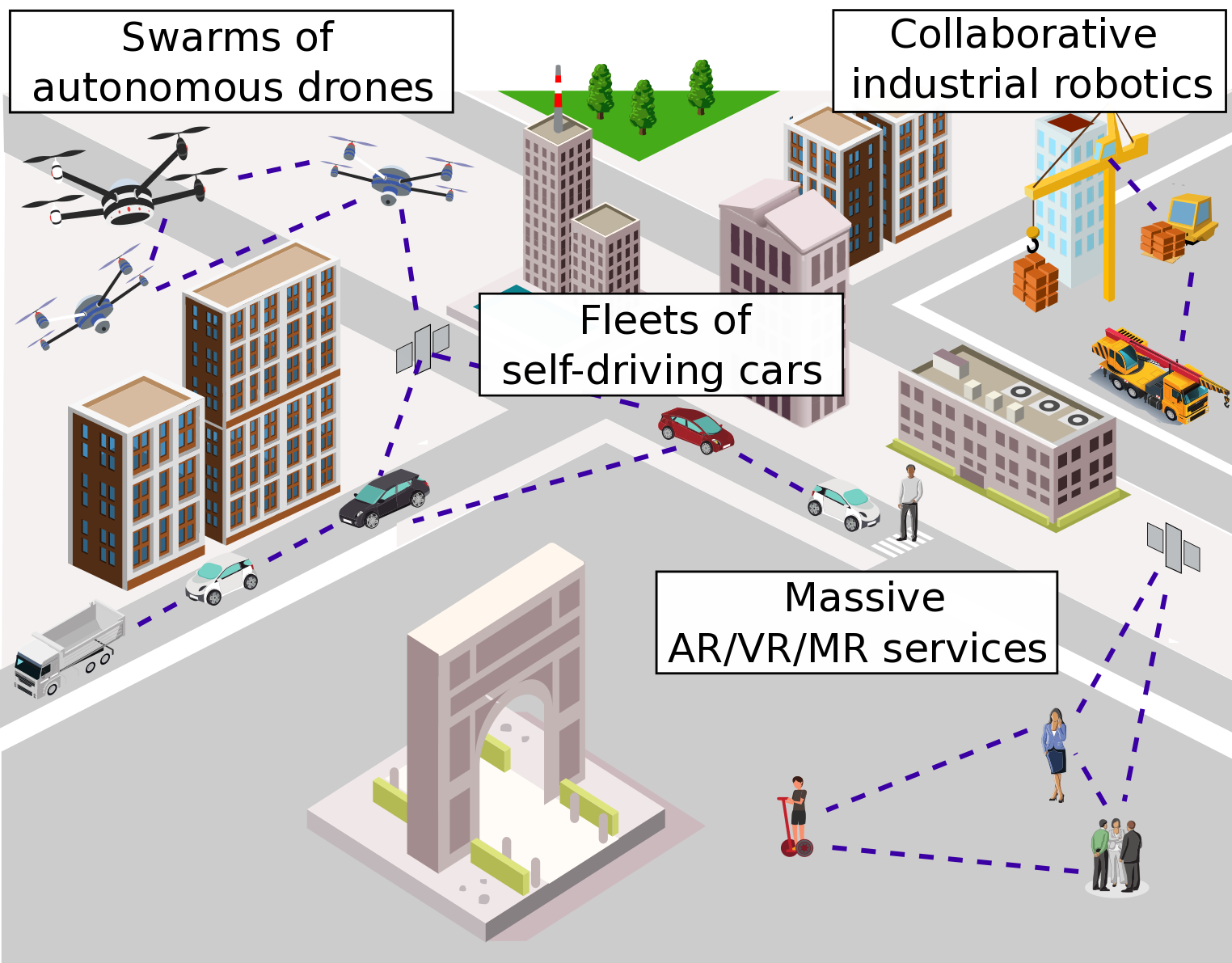}
    \caption{Envisaged advanced IoT applications and services.}
    \label{fig:applications}
\end{figure}

As a result, the SA 5G NR technology is expected to be at the foundation of the emerging 5G networks and facilitate their applications having stringent performance requirements. These include large fleets of connected and self-driving cars, swarms of autonomous drones, collaborative industrial robotics, as well as massive augmented, virtual, or mixed reality (AR/VR/MR) services (see Fig.~\ref{fig:applications}). Characteristic to all of these advanced use cases is unconstrained mobility of the devices as they exchange heavy latency- and reliability-sensitive data~\cite{Ericsson} between one another and with the proximate radio access infrastructure. Providing efficient support for these demanding \emph{`intelligent' IoT} applications is very different from operating mobile broadband or sensor-based services.

Despite significant recent developments in 5G radio access network (RAN) design, enabling cost-effective and scalable service provisioning for intelligent IoT devices that move unpredictably and potentially at high speeds requires a matching effort from the side of the network architecture~\cite{newJSAC}. In the past, cloud infrastructures were instrumental to maintain massive IoT deployments by providing with on-demand data processing and storage capabilities. More recently, untethered device mobility in next-generation IoT deployments aggressively pushed storage and computing resources toward dispersion. Along these lines, the edge computing paradigm has emerged to employ computation resources in the immediate vicinity of the IoT equipment for preliminary information processing and its localized storage.

However, relying on the edge nodes alone may become insufficient when dissimilar IoT services compete for their limited resources~\cite{7543455}, thereby imposing longer processing delays and reducing operational reliability. To alleviate these serious constraints, fog computing has been originally introduced by Cisco and nicknamed ``the cloud close to the ground''. In essence, fog architecture leverages end devices and near-device edge network infrastructure to achieve improved computation, communication, control, and storage capabilities as compared to legacy cloud-centric system design. To this aim, fog computing deploys virtualized cloud-like functionality closer to the end-user, by decisively extending the conventional paradigm and thus becoming broader than the typical notion of edge.

Enabling these lightweight cloud-like functions in close proximity to various moving devices, fog architecture permits the shifting of computationally-intensive services from the cloud to the edge~\cite{8016573} and can, therefore, serve them with direct `short-fat' wireless connections as opposed to the `long-thin' mobile cloud connections. In turn, this helps relieve the burden of fronthaul congestion as well as mitigate the overload of cloud data centers by facilitating more flexible and customizable performance of large-scale and dynamic layouts. In intelligent IoT, fog architectures are becoming vital not only because they are naturally suited for the very large numbers of moving devices, but also due to their inherent reliance on collective and location-based operation for enhanced system-level micro-management.

Indeed, future IoT envisions increasingly massive deployments of smarter entities, such as connected and autonomous vehicles, on the road and in the air, as well as moving robots and advanced AR/VR/MR gear. These capable devices are typically equipped with improved means for sensing, processing, storage, and communication as well as have more abundant on-board power sources to substantiate collaborative engagement. This includes but is not limited to collective processing and computing, distributed caching and storage, as well as cooperative resource management and control.

{
\color{black}
However, we argue that cost-effective support of large-scale intelligent IoT services at certain levels of reliability and on the move requires a dense geographical distribution of fog functionality, which we outline in this article and name \textit{dense moving fog}. We first review the key enablers for moving fog operation in Section~\ref{sec:enablers}. In Section~\ref{sec:multi_rat_iot}, we describe the dense moving fog functions by highlighting the advantages as well as outlining the challenges related to its conceptual implementation. Later, in Section~\ref{sec:numerical}, we present our case study for dense moving fog assistance in routine operations. Finally, we conclude this work by identifying future prospects on dense moving fog in Section~\ref{sec:conclusions}.
}

\section{Enablement of Fog Operation for Moving IoT}
\label{sec:enablers}

\subsection{Flexible Softwarized 5G Networking}

Intelligent IoT systems are envisioned to operate in highly-dynamic and heterogeneous environments. This is aggravated by the increasingly complex mobility patterns, including the unconstrained 3D mobility of drones and micro-scale mobility of wearable devices. Therefore, next-generation IoT devices need to thoroughly leverage all of the available proximate computing, connectivity, and caching options~\cite{7537178}, which goes beyond the conventional fixed topologies of 4G/4G+ networks. In response to these demands, an emerging capability of 5G system architecture as defined by 3GPP and \textcolor{black}{European Telecommunications Standards Institute} (ETSI) is its modular principle of the network function design.

Flexible operation is facilitated by the adoption of software-defined networking (SDN), which decouples the software-based control plane from the hardware-based data plane, thus allowing to adjust the behavior of network nodes in an automated and dynamic manner. Another important innovation is in the adoption of network function virtualization (NFV), which deploys crucial network functions as software components named virtual network functions (VNFs). Intelligent implementation of the SDN and NFV concepts in 5G networks leads to decisive benefits~\cite{8053453}, such as more efficient \textcolor{black}{and close-to-real-time} resource allocation as well as native integration of multiple radio access technologies (RATs), which results in more flexible data traffic management.

Softwarized 5G networking additionally introduces the concept of network slicing, which is essentially a virtual separation of heterogeneous data flows from one another by offering reservation of the network resources to a particular IoT application or service. In this case, an SDN-controlled 5G network can provide with a certain level of guarantees for the end-to-end on-time delivery of latency- and reliability-sensitive data traffic from the network nodes, thus offering a robust underlying platform for control, intelligence, computation, and data distribution in flexible fog operation (see Fig.~\ref{fig:network}).

\subsection{Multi-Hop Multi-Connectivity over mmWave}
As contemporary IoT devices are becoming more intelligent, `high-end' IoT solutions are already equipped with increasingly advanced sensors, including high-resolution visual and infrared \textcolor{black}{cameras, radars, lidars}, etc. Hence, the aggregate volume of data per hour of their operation can easily exceed hundreds of gigabytes~\cite{7786130}, thus calling for efficient mechanisms to exchange such information. Existing microwave technologies cannot provide with gigabit-per-second data rates with adequate latency and reliability guarantees. Therefore, utilization of the millimeter-wave (mmWave) band is envisioned as a key enabler for intelligent IoT services.

Wireless connectivity over the mmWave band holds the potential to unlock the much needed additional bandwidth and achieve close to noise-limited communication, as enabled by highly-directional transmissions of large-scale antenna arrays. Meanwhile, the narrower `pencil' beams may become occluded by most moderate-size objects, including buildings, large billboards, vehicle bodies, pedestrians, and many more.  The above challenges the reliability of data exchange over mmWave frequencies. As a result, the major standards developing organizations, including 3GPP and IEEE, are presently investing significant effort in improving the levels of reliability over mmWave (with e.g., 5G NR and IEEE 802.11ay technologies). Here, efficient mechanisms to maintain several simultaneous spatially-diverse 5G NR connections between the communicating nodes, termed multi-connectivity, are demanded to mitigate link blockage and achieve reliable mmWave networking.

However, in intelligent IoT scenarios where the communicating devices are highly mobile and the surrounding environment is dynamic~\cite{6815894}, multi-connectivity mechanisms are becoming vital, as the direct access to the static mmWave \textcolor{black}{radio access network} (RAN) or the intended communicating peer may be temporarily unavailable. Hence, `single-hop' multi-connectivity solutions can be insufficient and might be augmented with `multi-hop' functionality, which in turn benefits from the capability of a softwarized 5G network to support complex dynamic topologies. The establishment, maintenance, and timely update of dynamic `multi-hop' multi-connectivity links in high-speed mmWave communication is envisioned as one of the emerging enablers in collaborative fog-aided IoT systems (see Fig.~\ref{fig:network}).

\subsection{Learning and Artificial Intelligence in IoT}

The increased levels of dynamics and complexity in emerging network and access architectures call for fundamentally different decision-making capabilities to maintain radio connections, process and distribute data, as well as collaboratively manage the system. Beyond the distributed learning approaches, artificial intelligence (AI) solutions emerge to integrate all of the required fog functionalities~\cite{DBLP:journals/corr/ZhangDW16}. This vision is aggressively pushed forward by the advent of autonomous and self-driving vehicles, which may produce a disruption in intelligent transportation systems.

Existing solutions for autonomous cars rely on localized sensing and control. However, they face fundamental constraints in efficiency, reliability, and safety due to limited perception of local sensors. Recent accidents with autonomous vehicles confirm the importance of coordination between them. Due non-negligible latencies, the role of cloud-aided networking has been reduced to providing static or slowly varying support (e.g., traffic conditions and routes). To unleash the full potential of autonomous fleets, it is crucial to achieve cloud-integrated cooperative sensing and learning. However, this poses major challenges for computing and statistical inference with a need to develop hierarchical learning architectures, named federated learning, which will integrate distributed (local) and centralized learning.

\begin{figure}[!t]
    \centering
    \includegraphics[width=0.8\columnwidth]{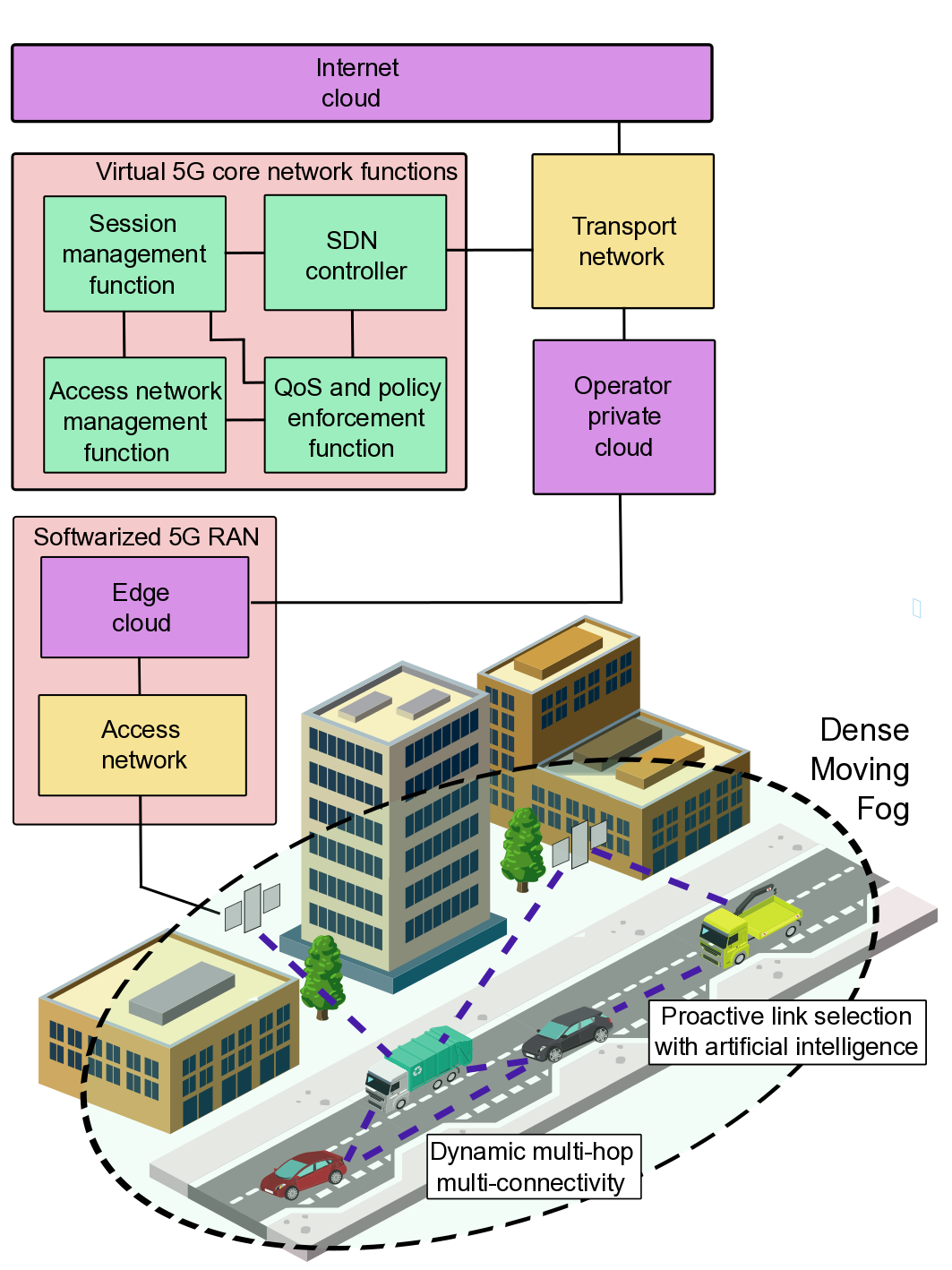}
    \caption{Functionality of dense moving fog for intelligent IoT applications.}
    \label{fig:network}
\end{figure}

Owing to high-rate mmWave connectivity and seamless cloud assistance, vehicles can cooperate to substantially enhance the accuracy of their localization and sensing (see Fig.~\ref{fig:network}). The benefit of a cloud-based learning solution is in its global perspective on cooperative sensing over cars. To this end, cloud-based cooperative fusion and vehicle-based local sensing can be integrated under a hierarchical learning architecture. The cloud-based solution can provide individual cars with adaptive side-information that reflects the global and cooperative view of the system. The vehicle-based solution can leverage the cloud assistance together with local sensor measurements to enhance the accuracy of its perception. Efficient data transfer techniques will need to be designed to connect these two inference operation modes.

\section{Dense Moving Fog and its Main Functionality}
\label{sec:multi_rat_iot}

\subsection{Computation}
Recently, fog architectures were envisioned to achieve rapid and affordable scaling by enabling computation capabilities flexibly along the entire ``cloud-to-things continuum''~\cite{7901470}. Accordingly, they help avoid resource contention at the edge of the network due to on-device data processing and cooperative radio resource management, further augmented with advanced communication, control, and storage capabilities. As a result, fog computing emerges as a unified end-to-end platform for a rapidly growing variety of dissimilar IoT services, which leverages the on-demand scalability of cloud resources as well as coordinates the involvement of geographically distributed edge and end devices. Hence, it has the potential to provide a rich set of fog computing functions for a large number of vertical IoT industries.

For intelligent IoT equipment with abundant processing and computation resources, dense fog seeks to realize their seamless integration with proximate edge equipment and remote cloud functions. This vision goes beyond treating the network edges and end devices as isolated computing platforms. Seamless integration of fleets and swarms of moving IoT entities into a dense fog enclave becomes a new distributed computing paradigm that improves scalability, extensibility, and compositionality of cloud-like services deployed closer to the network edge. In contrast to past fog-like considerations for static and low-power IoT modules (sensors, meters, actuators, etc.), more advanced capabilities of intelligent IoT devices (cars, drones, robots, etc.) make the energy costs of collaborative fog operation truly pay off.

{
\color{black}

\emph{\underline{Challenges in Computation for Dense Fog:}}

\subsubsection{Real-time task decomposition and load balancing} 
Owing to unconstrained mobility and extreme heterogeneity of the underlying computation substrate -- a unique feature of the dense moving fog -- efficient means are needed for on-the-fly processing, load balancing, and decomposition of computational tasks. Here, one may consider applying AI mechanisms, which are trained on the previous history of dense moving fog operations and thus able to promptly deliver a sufficiently appropriate task decomposition for the current system state. 

\subsubsection{Distributed computing over unreliable connectivity} 
Despite the very dense geographical distribution, catering for proximity to end devices in fog operation may not always be feasible due to possible lack of powerful computing modules, but also because of limited connectivity in critical conditions (e.g., intermittent and prone to failure wireless links). One of the possible approaches here is to determine the minimal sufficient levels of redundancy in task decomposition, so that the final outcome can be reconstructed even in cases where a part of the intermediate computation results is not available on time.
}

\subsection{Communication}
It is important to continuously provide adequate rates for data exchange in the considered scenarios, given that a connected car produces tens of megabytes of data per second, while an autonomous vehicle may generate up to a gigabyte per second~\cite{7498684}. Here, the dense moving fog can support the accelerated data traffic by heavily exploiting the directional high-rate communications over the mmWave bands. Dense moving fog can also provide novel ways for intelligent IoT devices to communicate with each other as well as with their proximate network infrastructure in the face of intermittent connectivity by utilizing the multi-hop multi-connectivity mechanisms, thus combining the advantages of centralized and ad-hoc network topologies into a unified solution.

{
\color{black}
\emph{\underline{Challenges in Communication for Dense Fog}:}

\subsubsection{(Ultra-)low-latency communication} 
Not limited to capacity demands, the emerging widely-deployed IoT applications may also require (ultra-)low latencies below a few tens of milliseconds: vehicle-to-everything communication, industrial and drone flight control, virtual reality and gaming services, etc. These latency-sensitive use cases may challenge the radio communication in dense moving fog. Fortunately, recent advances in embedded AI promise to `teach' the fog which job needs to be allocated to which resource to decrease the end-to-end delay and reduce the network loading. Here, delay-tolerant tasks may be pushed into the cloud, while time-sensitive operation can employ nearby fog devices, thus allowing the data to reside close to where it is being generated. 

\subsubsection{Reliable data exchange over opportunistic connectivity} 
To facilitate dynamic management of computing, networking, and storage functionalities, dense moving fog architecture needs to exercise real-time control along the continuum between the data centers and the end devices~\cite{8029686}. More flexible multi-hop and multi-connectivity solutions enabled over softwarized 5G radio networks~\cite{7020884} are thus envisioned to address this challenge. However, these mechanisms are currently at the early stages of their development and hence call for further research. This includes determining the optimal degree of multi-connectivity (the number of simultaneous links) in particular operating conditions of the dense moving fog. 
}

\subsection{Storage}

As fog infrastructures bring a plethora of cloud-like services closer to the end devices, efficient storage is crucial. Correspondingly, elastic memory capacity can be made available to various applications running on top of constrained IoT devices. Given that dimensioning of fog-aided operation is inherently flexible, the very large numbers of densely distributed and potentially mobile intelligent IoT entities may be integrated therein. Abundant storage space becomes accessible by the fog devices collectively with e.g., end nodes coalescing into ad-hoc capacity enclaves. As a result, multiple interconnected fog infrastructures that co-exist in space and time may serve as storage backup for each other by pooling various resources of the network edge, access, and end devices in proximity. 

{
\color{black}
\emph{\underline{Challenges in Storage for Dense Fog}:}

\subsubsection{Proactive storage selection procedures} 
To provide with flexible storage capabilities, dense moving fog has to enable informed and timely decisions on how to dynamically (re-)distribute the data among heterogeneous fog nodes. Here, proactive cell selection procedures, cooperative caching policies, and radio resource management strategies will be instrumental to address this challenge, achieve improved hit ratios at the edge, and thus avoid transferring massive data by reducing bandwidth consumption.

\subsubsection{Mobile big data analytics} 
Fog-enabled storage may benefit from mobile big data analysis, especially over densely distributed and increasingly heterogeneous data collection points. However, signaling overheads should be carefully controlled in this context, since the relatively frequent exchange of small-data packets (e.g., for traffic monitoring and logging purposes) may quickly deteriorate the available link budget. Hence, data collection and analysis have to be offered locally -- by utilizing end devices and/or edge-network infrastructure for more efficient micro-management. Augmenting data analytics with device-aided content sharing can further boost responsiveness and location awareness.
}

\subsection{Security}

Not limited to the above angles, fog infrastructures promise unique security-related opportunities. Massive and dense moving fog with already established dynamic chains of trust can act as a trusted authority for external devices and systems. Particularly, the moving fog may handle the responsibilities of a trusted computation platform, a certification authority, and a secure storage for short-lived sensitive information, among many others. Fog can also facilitate localized threat monitoring, detection, and protection for its nodes as well as offering powerful proximity-based authentication mechanisms by proxying the end devices for better identity verification.

{
\color{black}
\emph{\underline{Challenges in Security for Dense Fog}:}

\subsubsection{Secure operations in heterogeneous environment} 
The central concern here is that of heterogeneity: multiple potentially competing service providers and consumers are utilizing distributed and dissimilar resources across a diverse collection of hardware platforms in multi-tenant environments. Therefore, advanced authorization and authentication mechanisms will need to be coined, which will effectively leverage this heterogeneous medium and mediate between the fog entities~\cite{s17061421}. Fortunately, trusted execution environments supported by public-key infrastructures may become a suitable remedy for the above issues. Still, intelligent integration of hardware-assisted and software-centric security mechanisms remains an open research challenge for the envisaged dense moving fog.

\subsubsection{Dynamic adaptation of security measures} 
In contrast to the state-of-the-art systems primarily operating in known conditions, the prospective dense moving fog will have to handle volatile environments. Therefore, the employed security mechanisms have to continuously adapt to the current operating conditions. Facing this challenge, dense fog has to dynamically adjust its overall security level, which calls for designing new security protocols that will be ready to respond adequately to any security compromises without creating disruptions hampering safe and uninterrupted system operation.
}

\section{Example Use Case: Fog-Enabled Massive Fleet}
\label{sec:numerical}



\subsection{Scenario Description: Urban Fleet of Autonomous Cars}

\begin{figure}[!ht]
    \centering
    \includegraphics[width=0.95\columnwidth]{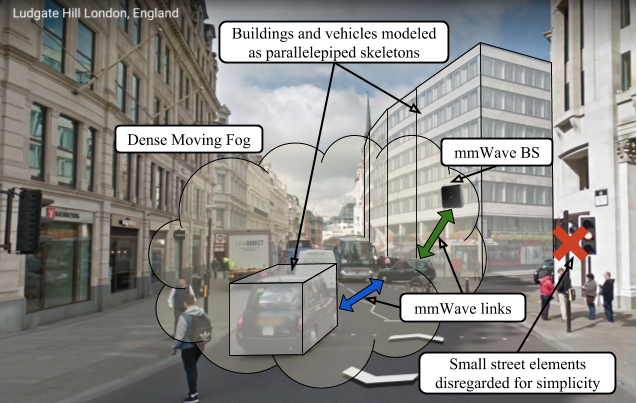}
    \caption{Part of simulation scenario overlaid with modeling considerations.}
    \label{fig:scenario}
\end{figure}

We address a typical urban deployment characteristic of the City of London, UK and bounded by Charterhouse St. to the North, Queen Victoria St. to the South, Moorgate to the East, and A201 to the West. Buildings are modeled as parallelepipeds with appropriate heights, while vehicles are represented as boxes of $4.8$\,m long, $1.8$\,m wide, and $1.4$\,m tall (see Fig.~\ref{fig:scenario}). The density of vehicles in the streets is estimated with the Google Street View. Each of the autonomous cars is equipped with mmWave radio transceivers, thus forming a connected fleet. The stationary mmWave infrastructure is deployed with the inter-site distance of $200$\,m. Each of the vehicles is further equipped with rich sensing capabilities. The rest of the modeling parameters are summarized in Table~\ref{tab:params}.

In our study, we specifically focus on the capability of a connected fleet to avoid collisions with one of the major threats for autonomous vehicles in urban environments -- jaywalking -- when a pedestrian is crossing the road illegally. Hence, we mimic these events as a random process inside the area of interest with the intensity of $1$ cross per minute. The location of a jaywalking attempt is randomized over the area, while the speed of a jaywalking pedestrian is set to $10$\,km/h. We demonstrate that while a single car has certain chances to miss a running person, the fog-aided fleet of self-driving vehicles is capable of detecting this dangerous event and responding to it with appropriate collective reaction even under stringent delay constraints. 

Collective sensing of the area by multiple autonomous vehicles and collaborative data processing of thus sensed data by both the vehicles and the road infrastructure becomes particularly beneficial for the considered scenario. We employ our in-house system-level simulation framework to quantify the described use case and obtain the first-order performance results. The said framework is implemented in Python and follows the latest 3GPP guidelines for mmWave radio propagation and physical layer modeling. 

\textcolor{black}{Our tool is a time-driven simulator with the step of $0.01$\,s. To improve the accuracy of the output results, all of the intermediate output has been averaged over $100$ independent simulation rounds, which corresponds to approximately $17$\,hours of real-time operation (each of the rounds models $10$\,min of real time).} The corresponding latencies of data exchange across the connected fleet are carefully taken into account together with other important fog-based enhancements pertaining to moving IoT devices.

\subsection{Support of Fog Enhancements in Moving IoT}

\subsubsection{Flexible Softwarized 5G Networking}

In our scenario, the utilization of softwarized 5G networking allows for constructing flexible temporarily-optimal network configurations without significant costs. Further, we require complete liquidization of spectrum resources at both RAN and \textcolor{black}{core network} (CN) sides. The amount of spectrum available for a certain link is modeled as a continuous variable, which is infinitely and dynamically (re-)divisible subject to the current number of simultaneous data streams as well as their rate demands. Here, the virtual network `oracle' observes the status of all radio links to dynamically adjust the connectivity and routing maps for maximizing the target metrics of interest.

\begin{table}[!b]
\caption{Example modeling parameters.}
\label{tab:params}
\begin{center}
\begin{tabular}{p{0.3\columnwidth}|p{0.65\columnwidth}}
\hline
\multicolumn{2}{c}{\textbf{Deployment}}\\
\hline
Area of interest & London City, UK\\
\hline
Building models & From the map with real heights (see Fig.~\ref{fig:scenario})\\
\hline
\multicolumn{2}{c}{\textbf{mmWave radio}}\\
\hline
Carrier frequency & $28$\,GHz\\
\hline
Bandwidth & $500$\,MHz (allocated for fog with network slicing)\\
\hline
Propagation model & 3GPP \textcolor{black}{urban microcell} (UMi) -- street canyon\\
\hline
Effect of buildings & 3GPP LoS $\rightarrow$ 3GPP nLoS\\
\hline
Effect of vehicles & Non-blocked $\rightarrow$ blocked ($20$\,dB degradation)\\
\hline
ISD of stationary BSs & $200$\,m\\
\hline
BS sectorization & $3$ sectors per site with $102\degree$ downtilting\\
\hline
BS transmit power & $35$\,dBm\\
\hline
BS height & $10$\,m\\
\hline
BS computing perf.& $3$\,TFLOPS \textcolor{black}{($10^{12}$ floating-point operations per second)}\\
\hline
Car-cell transmit power & $20$\,dBm (windshield transceiver location)\\
\hline
\multicolumn{2}{c}{\textbf{Vehicles}}\\
\hline
Body models & Parallelepiped $4.8$\,m $\times$ $1.8$\,m $\times$ $1.4$\,m\\
\hline
Driving speed & $40$\,km/h (constant) \\
\hline
Mobility pattern & Manhattan grid on all streets\\
\hline
Computing perf.& $5$\,TFLOPS\\
\hline
Radar sensing range & $50$\,m\\
\hline
Radar cycle duration & $66$\,ms\\
\hline
\multicolumn{2}{c}{\textbf{Pedestrians}}\\
\hline
Jaywalking speed & $10$\,km/h (constant)\\
\hline
Jaywalking intensity & $1$\,per min for scenario\\
\hline
\end{tabular}
\end{center}
\end{table}

\subsubsection{Multi-Hop Multi-Connectivity over mmWave}

Our scenario assumes that all of the mmWave transceivers support multi-connectivity of degree $3$ -- not more than three simultaneous mmWave links -- which is a compromise between session continuity and implementation complexity. Due to network softwarization, connected vehicles can support multi-hop communication, both as initiators and as sink nodes, while assisting the other cars in data forwarding. The maximum number of hops is not limited at the network layer but is rather determined by the application-specific requirements (e.g., latency constraints). A node is assumed to continuously follow the commands received from the `oracle' as well as update its active and backup links according to these instructions.

\subsubsection{Learning and Artificial Intelligence in IoT}

A major challenge in urban mmWave connectivity is random link quality drops due to radio signal blockage by an obstacle. While they can be mitigated with softwarized RAN and multi-connectivity features, these enhancements remain only a partial solution. Fortunately, mobility in vehicular environments is more predictable than directions of human travel\textcolor{black}{~\cite{7786130}}. Further, self-driving vehicles are equipped with numerous sensors \textcolor{black}{(radars, lidars, cameras, etc.) to estimate their positions} as well as locations and relative velocity of the nearby objects. \textcolor{black}{Smart cars may also utilize microwave vehicular communication systems -- IEEE 802.11p, dedicated short range communications (DSRC), etc. -- for assisted node discovery or mmWave medium access control tasks.} If supplied with appropriate intelligence, an autonomous vehicle can thus predict a blockage event of its mmWave link(s) in advance. We model such proactive connectivity (re)selection, where a car may anticipate a link failure and reconnect to another more reliable alternative.


\subsection{Understanding Benefits of Dense Moving Fog}

\begin{figure}[!ht]
    \centering
    \includegraphics[width=0.9\columnwidth]{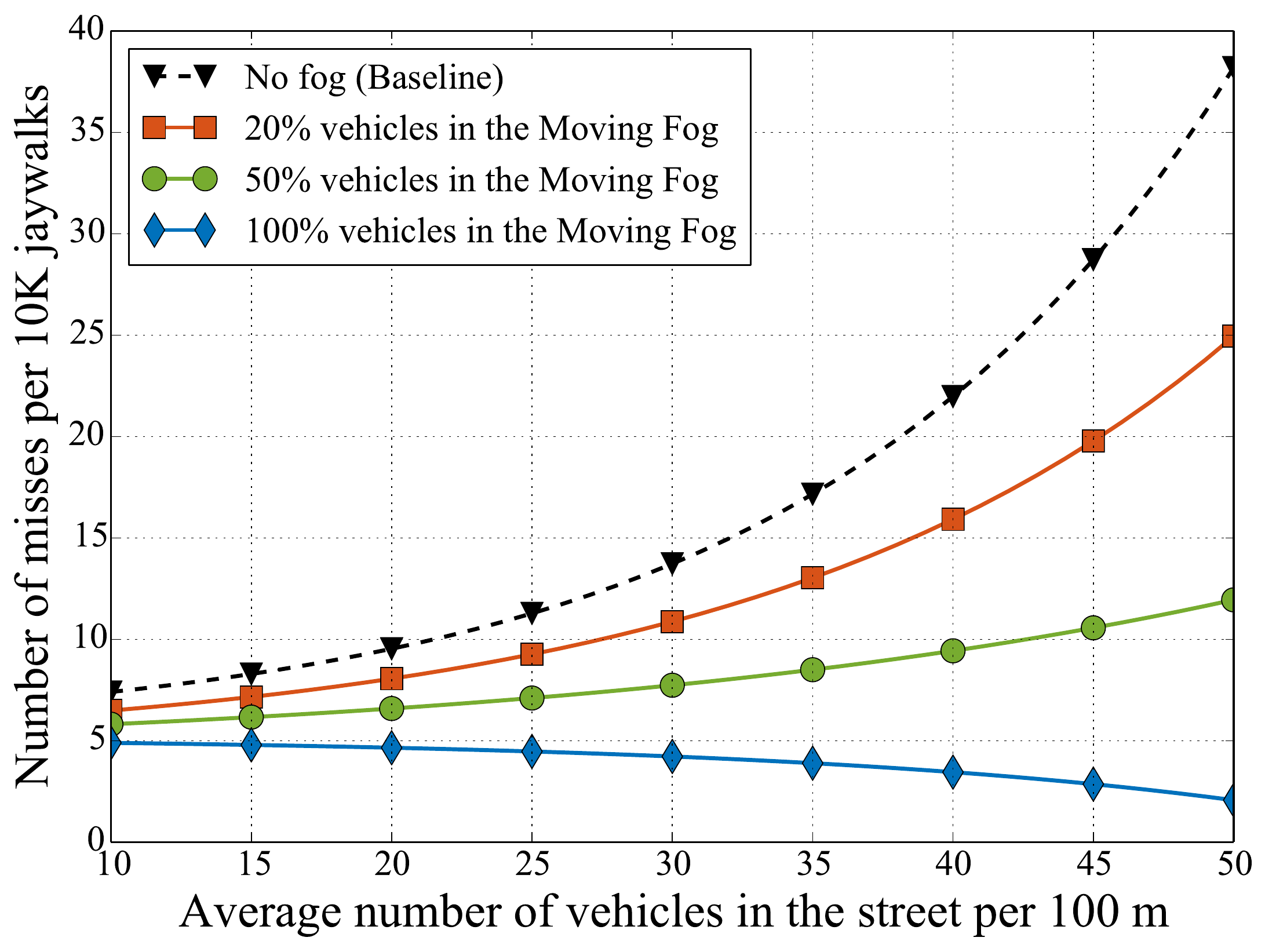}
    \caption{Impact of dense moving fog on detection of jaywalking pedestrians.}
    \label{fig:plot1}
\end{figure}

Here, we summarize our first-order analysis of the benefits brought by dense moving fog to a connected fleet. We begin with Fig.~\ref{fig:plot1} that illustrates the number of misses in jaywalking detection as a function of the density of vehicles in the streets ($4$ lanes assumed). As can be observed, higher densities of cars impact the jaywalking detection negatively in the baseline scenario, since vehicle bodies block the sensory capability to detect the event on time: pedestrians suddenly appear from behind the other vehicle. Once the dense moving fog becomes operational, this negative effect of vehicle densification is complemented by the positive trend of the increased sensing capability in a collaborating fleet as well as by collective data processing. 

To this aim, Fig.~\ref{fig:plot1} confirms that even a small fraction of vehicles in the fog (e.g., $20\%$) allows for a significant -- over $50\%$ -- reduction in the miss rate ($24.7$ vs. $38.6$ for $50$ vehicles per $100$\,m). For higher penetration of the moving fog functionality, this positive effect starts dominating, since the connected fleet can assess the same spot from different angles, predict the jaywalk trajectory, and warn proximate cars about the threat. Here, two important observations are made as a result: (i) the moving fog is more beneficial at higher densities of the involved entities as well as (ii) it enables assistance in real-world applications and services far beyond the vertical \textcolor{black}{information and communication technologies} (ICT) use cases, e.g., improved chances of on-time jaywalking detection.

\begin{figure}[!ht]
    \centering
    \includegraphics[width=0.95\columnwidth]{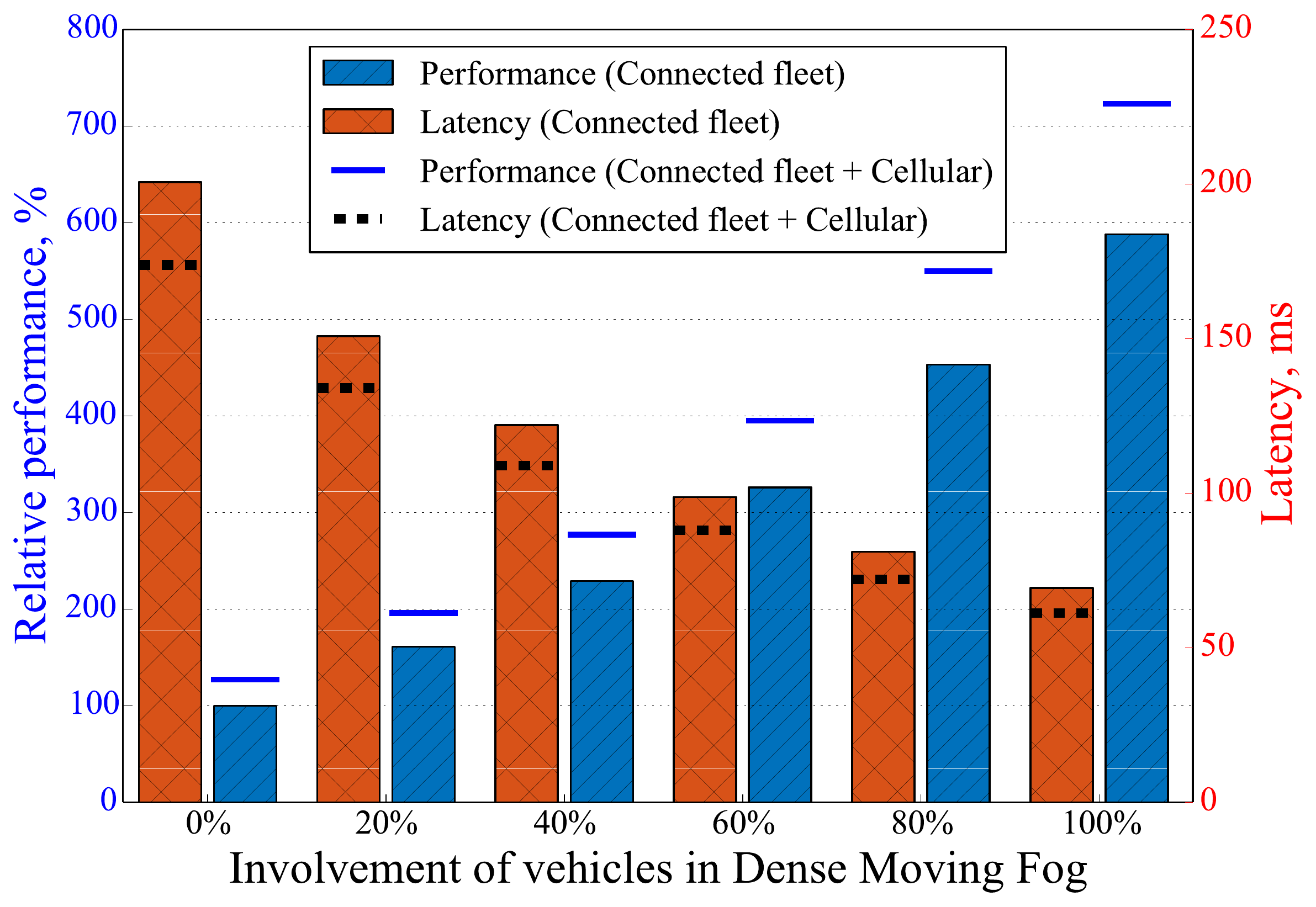}
    \caption{Impact of dense moving fog on collaborative data processing.}
    \label{fig:plot2}
\end{figure}

Continuing with Fig.~\ref{fig:plot2}, we generalize the above point use case to investigate the mechanics of dense moving fog in enabling the aforementioned gains. We first compare the aggregate benefit of the vehicular fog vs. the performance of a single autonomous car (set as $100\%$ for reference). One can observe that a random vehicle in the fog may receive a boost of up to $6$\,times in its on-time data processing rate by sharing the computational load with other vehicles in proximity. In particular, we consider only those vehicles that respond to a computation offloading request within $50$\,ms (a car moves only $0.5$\,m during that time, which ensures its relevant response). Similar trends are observed for other threshold values.

The connected fleet may further benefit from offloading its tasks onto the edge RAN infrastructure. Hence, we assume that each mmWave \textcolor{black}{base station} (BS) shares its computational resource equally between all served vehicles. Higher involvement factors yield additional improvements since a larger and denser fog is more likely to interconnect across multiple underloaded BSs. Better performance naturally leads to lower processing times. More specifically, a fixed-size job of $10^{12}$ floating-point operations ($1$ TFLOP) will be computed by a standalone vehicle within $200$\,ms, while the dense fog (assuming efficient parallelization) is able to decrease this value down to $69$\,ms (by $3$\,times). Then, $\approx$$15\%$ more is gained if the network infrastructure also becomes involved in collaborative data processing.

\section{Future Prospects of Dense Fog Operation}
\label{sec:conclusions}

In conclusion, our numerical results confirm the benefits of the moving fog infrastructure not only for a particular application (e.g., connected car fleet) but also in terms of collaborative data processing across many practical use cases along the lines of vehicular and airborne fog computing. \textcolor{black}{The benefits will grow even further if the data sensed by a dense moving fog are complemented with those produced by the networking and/or road infrastructure. The latter requires resolving additional challenges related to matching the formats of heterogeneous data streams. This becomes a promising research direction to extend our present work.}

The advantages of dense fog stretch above and beyond rapid and affordable scaling: it will unlock emerging IoT services and disruptive business models as well as help accelerate the roll-outs of new products by creating a more open marketplace. As cloud and fog architectures converge within an integrated end-to-end platform, numerous angles of this rapidly materializing multi-faceted vision will need to be aligned in future studies on the subject, which includes performance, reliability, safety, business, and regulatory issues.

\balance
\bibliographystyle{ieeetr}
\bibliography{references_short}

\section*{Authors' Biographies}

\textbf{Sergey Andreev} (sergey.andreev@tut.fi) is an assistant professor in the Laboratory of Electronics and Communications Engineering at Tampere University of Technology, Finland. He received the Specialist degree (2006) and the Cand.Sc. degree (2009) both from St. Petersburg SUAI, as well as the Ph.D. degree (2012) from TUT. Since 2018, he has also been a Visiting Senior Research Fellow with the Centre for Telecommunications Research, King's College London. Sergey (co-)authored more than 150 published research works on wireless communications, energy efficiency, and heterogeneous networking.

\textbf{Vitaly Petrov} (vitaly.petrov@tut.fi) is a Ph.D. candidate at the Laboratory of Electronics and Communications Engineering at Tampere University of Technology (TUT), Finland. He received the Specialist degree (2011) from SUAI University, St. Petersburg, Russia and the M.Sc. degree (2014) from TUT. He was a Visiting Scholar with Georgia Institute of Technology, Atlanta, USA, in 2014. Vitaly (co-)authored more than 30 published research works on millimeter-wave/terahertz band communications, Internet-of-Things, nanonetworks, cryptology, and network security.

\textbf{Kaibin Huang} (huangkb@eee.hku.hk) is an assistant professor in the Department of EEE at the University of Hong Kong. He is an editor for IEEE Transactions on Wireless Communications and also IEEE Transactions on Green Communications and Networking. He received a Best Paper Award from IEEE GLOBECOM 2006 and an IEEE Communications Society Asia Pacific Outstanding Paper Award in 2015.

\textbf{Maria Lema} (maria.lema\_rosas@kcl.ac.uk) holds a Ph.D. in Wireless Communications and an MSc. in Telecommunication Engineering \& Management from UPC. Currently responsible for technical operations in the 5G Testbed and Project Management for multiple activities related to Technology Transformation of industry verticals with 5G. She is involved in the definition of applications for 5G, working together with various industries to identify the main requirements and challenges to successfully bring 5G to market.

\textbf{Mischa Dohler} (mischa.dohler@kcl.ac.uk) is full professor in Wireless Communications at King's College London, driving cross-disciplinary research and innovation in technology, sciences, and arts. He is a Fellow of the IEEE, the Royal Academy of Engineering, the Royal Society of Arts (RSA), the Institution of Engineering and Technology (IET); and a Distinguished Member of Harvard Square Leaders Excellence. He is a serial entrepreneur; composer \& pianist with 5 albums on Spotify/iTunes; and fluent in 6 languages.

\end{document}